

\magnification=\magstep1       	
\font\bigbold=cmbx10 scaled 1200
\def\Bbb#1{{\cal #1}}          	
\let\frak=\cal			

\newcount\EQNO      \EQNO=0
\newcount\FIGNO     \FIGNO=0
\newcount\REFNO     \REFNO=0
\newcount\SECNO     \SECNO=0
\newcount\SUBSECNO  \SUBSECNO=0
\newcount\FOOTNO    \FOOTNO=0
\newbox\FIGBOX      \setbox\FIGBOX=\vbox{}
\newbox\REFBOX      \setbox\REFBOX=\vbox{}
\newbox\RefBoxOne   \setbox\RefBoxOne=\vbox{}

\expandafter\ifx\csname normal\endcsname\relax\def\normal{\null}\fi

\def\Eqno{\global\advance\EQNO by 1 \eqno(\the\EQNO)%
    \gdef\label##1{\xdef##1{\nobreak(\the\EQNO)}}}
\def\Fig#1{\global\advance\FIGNO by 1 Figure~\the\FIGNO%
    \global\setbox\FIGBOX=\vbox{\unvcopy\FIGBOX
      \narrower\smallskip\item{\bf Figure \the\FIGNO~~}#1}}
\def\Ref#1{\global\advance\REFNO by 1 \nobreak[\the\REFNO]%
    \global\setbox\REFBOX=\vbox{\unvcopy\REFBOX\normal
      \smallskip\item{\the\REFNO .~}#1}%
    \gdef\label##1{\xdef##1{\nobreak[\the\REFNO]}}}
\def\Section#1{\SUBSECNO=0\advance\SECNO by 1
    \bigskip\leftline{\bf \the\SECNO .\ #1}\nobreak}
\def\Subsection#1{\advance\SUBSECNO by 1
    \medskip\leftline{\bf \ifcase\SUBSECNO\or
    a\or b\or c\or d\or e\or f\or g\or h\or i\or j\or k\or l\or m\or n\fi
    )\ #1}\nobreak}
\def\Footnote#1{\global\advance\FOOTNO by 1
    \footnote{\nobreak$\>\!{}^{\the\FOOTNO}\>\!$}{#1}}
\def\SameFootnote{$\>\!{}^{\the\FOOTNO}\>\!$}

\def\References{\bigskip\centerline{\bf REFERENCES}
                \smallskip\copy\REFBOX}
\def\NewRefPage{\setbox\RefBoxOne=\vbox{\unvcopy\REFBOX}%
		\setbox\REFBOX=\vbox{}%
		\def\References{\bigskip\centerline{\bf REFERENCES}
                		\nobreak\smallskip\nobreak\copy\RefBoxOne
				\vfill\eject
				\smallskip\copy\REFBOX}%
		\def\NewRefPage{}}






\def\Fig#1{\global\advance\FIGNO by 1 Figure~\the\FIGNO: {#1}%
    \global\setbox\FIGBOX=\vbox{\unvcopy\FIGBOX
      \narrower\smallskip\item{\bf Figure \the\FIGNO~~}#1}}

\newcount\ITEMNO     \ITEMNO=0

\def\itemno{\global\advance\ITEMNO by 1 \the\ITEMNO}


\long\def\Theorem#1#2{\medbreak\noindent{\bf Theorem \itemno
\enspace}{\sl#1}\par\medbreak {\medbreak\narrower {\it Proof:}
{#2}\hfill{$\spadesuit$}\medbreak\smallskip}}

\def\Theoremwoproof#1{\medbreak\noindent{\bf Theorem \itemno
\enspace}{\sl#1}\par\medbreak}

\def\Definition#1{\medbreak\noindent{\bf Definition \itemno
\enspace}{\sl#1}\par\medbreak}

\def\Proposition#1#2{\medbreak\noindent{\bf Proposition \itemno
\enspace}{\sl#1}\par\medbreak {\medbreak\narrower {\it Proof:} #2\
$\spadesuit$\medbreak\smallskip}}

\nopagenumbers			
\headline={\ifnum\pageno=1{\hss}\else{\hss\rm -~\folio~- \hss}\fi}


\def\cross{\times}



\def\gamu#1#2#3{\gamma^{#1}_{\ #2#3}}


\def\frac#1#2{{{#1}\over {#2}}}
\def\fpartial#1#2#3#4{\frac{\partial #1^{#2}}{\partial
#3^{#4}}}

\def\at#1{\lower.8ex\hbox{${\big\vert _{#1}}$}}

\def\comma#1{{_{,#1}}}
\def\starry#1{{_{*#1}}}
\def\pel#1{{\rm P}_i}
\def\dt{\frac{\partial}{\partial t}}

\def\dxi{\frac{\partial}{\partial x^i}}

\def\partiali{\partial_i}
\def\partialj{\partial_j}
\def\partialk{\partial_k}
\def\partiall{\partial_l}

\def\partialt{\partial_t}

\def\lstar#1{\hbox{\it
\$}_{\!\!*\,\,\lower2pt\hbox{$\!\!\scriptscriptstyle#1$}}}

\def\del#1{{\nabla_{#1}}}

\def\Sdel{{\nabla_{\!\!*}}}
\def\sdel#1{{\nabla_{\!\!*#1}}}


\def\rperpu#1#2#3#4{{^\perp\! R}^{#1}_{\ #2#3#4}}

\def\rperp{{^\perp\!R}}

\def\zelmanovu#1#2#3#4{Z^{#1}_{\ #2#3#4}}

\def\D{{\cal D}}				

\def\generalTensor#1#2#3#4#5{{#1^{#2 \dots #3}_{\phantom{#2 \dots #3}#4\dots
#5}}}

\def\sqr#1#2{{\vbox{\hrule height.#2pt
              \hbox{\vrule width.#2pt height#1pt \kern#1pt
                   \vrule width.#2pt}\hrule height.#2pt}}}


\def\hij{h_{ij}}

\def\metric#1#2{\left< #1 , #2 \right>}

\def\sbracket#1#2{\left[ #1 , #2 \right]_*}

\def\p1tensors{{\cal P}\! T_1}
\def\ptensors#1#2{{\cal T}^{#1}_{#2}}



\def\vfs{\raise3pt\hbox{$\chi$}(\M)}
\def\pvfs{\raise3pt\hbox{$\chi_{_*}$}(\Sigma)}
\def\pfcns{{\frak F}_{\!p}(\Sigma)}	



\def\j1pi{J^1_{\pi}}

\def\hforms{\bigwedge\,\raise6pt\hbox{$\!\!\scriptstyle
1$}\lower6pt\hbox{$\!\!\scriptstyle 0$} \pi}

\def\M{{\cal M}} 		

\def\at#1{{\Big\vert_{#1}}}
\def\pfcns{{\frak F}_*(\Sigma)}
\def\dt{\partialt}
\def\ip#1{i_{\scriptscriptstyle #1}}

\def\BR{{\Bbb R}}


\def\today{\number\day\space\ifcase\month\or
  January\or February\or March\or April\or May\or June\or
  July\or August\or September\or October\or November\or December\fi
  \space\number\year}
\rightline{5 July 1994}
\rightline{gr-qc/9407012}
\bigskip\bigskip

\null\bigskip\bigskip\bigskip
\centerline{\bigbold PARAMETRIC MANIFOLDS II: Intrinsic Approach}
\bigskip

\centerline{Stuart Boersma}
\centerline{\it Department of Mathematics, Oregon State University,
		Corvallis, OR  97331, USA
\Footnote{Present address: Division of Mathematics and Computer Science,
Alfred University, Alfred, NY  14802}
}
\centerline{\tt boersma@math.orst.edu}
\medskip
\centerline{Tevian Dray}
\centerline{\it Department of Mathematics, Oregon State University,
		Corvallis, OR  97331, USA}
\centerline{\tt tevian@math.orst.edu}

\bigskip\bigskip\bigskip\bigskip
\centerline{\bf ABSTRACT}
\midinsert

\narrower\narrower\noindent
A {\it parametric manifold} is a manifold on which all tensor fields depend on
an additional parameter, such as time, together with a {\it parametric
structure}, namely a given (parametric) 1-form field.  Such a manifold admits
natural generalizations of Lie differentiation, exterior differentiation, and
covariant differentiation, all based on a nonstandard action of vector fields
on functions.  There is a new geometric object, called the {\it deficiency},
which behaves much like torsion, and which measures whether a parametric
manifold can be viewed as a 1-parameter family of orthogonal hypersurfaces.

\endinsert
\vfill
\eject

\Section{Introduction}

It is often useful to project the geometric structure of a manifold onto an
embedded hypersurface.  This leads to the well-known Gauss-Codazzi formalism,
which relates the projected geometry of the hypersurface to the original
manifold.  Initial value problems are often posed in this setting, with a
1-parameter family of embedded hypersurfaces being used to describe the
evolution.  Identifying these hypersurfaces leads to the interpretation of
tensor fields in the original manifold as 1-parameter families of tensor
fields on a given hypersurface.  This is the beginnings of a theory of {\it
parametric manifolds}.

We recently generalized the Gauss-Codazzi formalism from the setting just
described to the case where the manifold is foliated by the integral curves of
a (suitably regular) vector field, but where these curves are {\bf not}
assumed to be hypersurface orthogonal
\Ref{Stuart Boersma and Tevian Dray,
{\it Parametric Manifolds I: Extrinsic Approach},
J. Math.\ Phys.\ (submitted).}\label\PaperI
.{}
We will refer to this as the {\it extrinsic} approach to parametric manifolds.
This results in a picture of a parametric manifold which is now the {\it
manifold of orbits} of the given curves, on which there are 1-parameter
families of tensor fields.

However, there are implicit properties which such parametric manifolds inherit
from the original manifold.  Notable among these is the behavior under {\it
reparameterizations}, which consist of relabelling the parameter along the
given curves, and which are hence a special class of coordinate
transformations in the original manifold.

We show here that parametric manifolds can be be defined completely {\it
intrinsically}, without reference to an ``original manifold''.  The key idea
is to generalize the action of vector fields on functions in way reminiscent
of the notion of horizontal lift in a fibre bundle.  This naturally leads to
generalizations of Lie differentiation, exterior differentiation, and
covariant differentiation.  These derivative operators reproduce intrinsically
the corresponding projected operators obtained in our earlier extrinsic
approach.

The geometry of parametric manifolds is ``almost a fibre bundle'', and as such
may provide the groundwork for a generalization of Yang-Mills theory.

We start by defining parametric manifolds in Section 2.  We then introduce
parametric exterior differentiation in Section 3, which allows us to define
the all-important notion of {\it deficiency}, which measures whether a
parametric manifold can be viewed as a 1-parameter family of orthogonal
hypersurfaces.  In Section 4, we then use the deficiency to define a parametric
bracket, and hence a parametric Lie derivative.  Parametric connections are
discussed in Section 5, including their associated (generalized) torsion and
curvature.  Finally, in Section 6, we discuss our results.


\Section{Parametric Functions and Vector Fields}

Consider a smooth manifold $\Sigma$.  We wish to consider 1-parameter families
of tensor fields on $\Sigma$, parameterized by a parameter $t$.  Since the
particular choice of parameter should not be important, we first need to
describe how to change the parameterization.
\Definition{A reparameterization of $\Sigma$ is an assignment
$$s = t + F(p)$$ for $p\in\Sigma$, $s,t\in\BR$, and $F:\Sigma\to\BR$.}
A {\it parametric structure} on $\Sigma$ is a preferred 1-parameter family of
1-forms $\omega(t)$ on $\Sigma$ which behaves as follows under a
reparameterization:
$$\hat\omega(s) = \omega(t) - dF\Eqno$$\label\ReparProp
{\it i.e.}\ $\omega(t)$ transforms to $\hat\omega(s)$ under a
reparameterization.

We can now start to consider parametric objects on $\Sigma$.
\Definition {A parametric function on $\Sigma$ is a mapping
$f:\Sigma\cross\BR\to\BR$.  Denote the collection of such mappings by
$\pfcns$.}

Given a parametric function $f\in\pfcns$, for a fixed $t\in\BR$ $f$ can be
considered as a function from $\Sigma$ to $\BR$.  Denote this function by
$f_t$.  Thus $f_t\in {\frak F}(\Sigma)$, the ring of functions on $\Sigma$,
and can be acted on by tangent vectors of $\Sigma$.

\Proposition{The action of $\dt$ on parametric functions is a covariant
operation.}
{Under a coordinate transformation of $\Sigma$, the operator $\dt$ remains
unaffected.  This is because the parameter $t$ is not a coordinate and, hence,
any coordinate transformation of $\Sigma$ must be independent of $t$.
Therefore, $\fpartial f{}t{}\at{(p,t_0)}$ does not depend on
the choice of coordinates for $p\in\Sigma$.  Furthermore, under a
reparameterization $s=t+F(p)$, $\fpartial f{}s{} = \fpartial f{}t{}.$}

While tangent vector fields do not act uniquely on parametric functions,
1-parameter families of tangent vector fields do.  These 1-parameter
families of vector fields, called parametric vector fields, will act on
parametric functions in a way reminiscent of the action of projected
vector fields.

\Definition{A parametric vector field is a smooth mapping
$X:\Sigma\cross\BR\to T\Sigma$ such that for each $p\in\Sigma$, $X(p,t) \in
T_p\Sigma$ for all $t\in\BR$.
Let $\pvfs$ represent the collection of smooth parametric vector fields
defined on $\Sigma$.}

For a fixed $t$, let $X_t:\Sigma\to T\Sigma$ denote the obvious tangent vector
field.  We define the action of a parametric vector field on a parametric
function as follows:
$$Xf(p,t) = X_tf_t(p) + \omega(t)\left(X_t\right)\frac{\partial f}{\partial t}
(p).$$
Suppressing the point $p$, we can write this action as
$$X(f) = X_t f_t + \omega(X_t) {\partial f \over \partial t}
  .\Eqno$$\label\PVFAction

\Theorem{$X(f)$ is invariant under reparameterizations and coordinate
transformations.}
{Consider coordinates $\lbrace x^i \rbrace$ and a parameter $t$.  Writing
$\omega=:M_i\,dx^i$, we have that
$$X(f) = X^i \left(\fpartial {f_t}{}xi + M_i \fpartial f{}t{}\right).$$
Under a reparameterization $s=t+F(p)$, the components of $\omega$ transform
according to equation \ReparProp .  Denote the parametric structure $\omega$
under this new parameterization by $\hat\omega$.  Thus,
$$\eqalign{\hat\omega &= \hat M_i\, dx^i\cr
	&= \left(M_i - \fpartial F{}xi \right) dx^i\cr
	&= \omega - dF.\cr}$$
Although $\fpartial f{}s{} = \fpartial f{}t{}$, we must be careful computing
$\fpartial f{}xi.$  Using the notation introduced above, let $f_t :
\Sigma\to\BR$ and let $\hat f_s(p) = f(p,s) = f(p,t+F(p))$ denote its
reparameterization.  Then
$$\eqalign{\fpartial {\hat f_s}{}xi &= \fpartial f{}xj \fpartial xjxi +
\fpartial
f{}s{} \fpartial s{}xi\cr
	&=\fpartial f{}xi + \fpartial f{}t{}\fpartial t{}s{} \fpartial
F{}xi\cr
	&= \fpartial {f_t}{}xi
		+ \fpartial F{}xi {\partial f \over \partial t}.\cr}$$
Therefore,
$$\eqalign{X(f) &= X^i \left(\fpartial {f_t}{}xi
		+ M_i \fpartial f{}t{}\right)\cr
	&= X^i\left( \fpartial {f_t}{}xi + \left( \hat M_i
		+ \fpartial F{}xi\right)\fpartial f{}t{}\right)\cr
	&=X^i\left(\left(\fpartial {\hat f_s}{}xi
		- \fpartial F{}xi \fpartial f{}t{}\right)
		+ \left (\hat M_i + \fpartial F{}xi\right)
			\fpartial f{}t{}\right)\cr
	&= X^i \left(\fpartial {\hat f_s}{}xi
		+ \hat M_i \fpartial f{}s{}\right)\cr}$$
which is the expression for $X(f)$ with respect to the parameter $s$, showing
that $X(f)$ is invariant under a reparameterization.
If we consider a coordinate transformation of $\Sigma$, $X_t$ and $\dxi$ will
transform as usual, guaranteeing that $X_t(f_t)$ is independent of the choice
of coordinates.  Since $\omega$ and $\dt$ are unaffected, $X(f)$ remains
invariant under a coordinate transformation of $\Sigma$.}

\Theorem{Parametric vector fields are derivations on the ring $\pfcns$.
That is,
\item{\it i.} $X(rf + sg) = rX(f) + sX(g)$ and
\item{\it ii.} X(fg) = fX(g) + gX(f)
for all $r,s\in\BR$ and $f,g\in \pfcns$.}
{This follows directly from \PVFAction\ since $X_t$ and $\dt$ are derivations.
}

Parametric vector fields have a very nice representation in terms of a local
coordinate system, $\{ x^i\}$.  Since a parametric vector field is just a
family of tangent vector fields, we may write
 $$X=X^i\frac{\partial}{\partial
x^i} = X^i\partiali$$
as usual, where we let the functions $X^i$ depend on the parameter.  That is,
the $X^i$ are parametric functions on $\Sigma$.  In terms of this
representation we may write out the action of parametric vector fields on
parametric functions
$$\eqalign{X(f) &=X_t(f_t) + \omega(t)(X_t)\dot f\cr
	&=X^i \Big( f\comma i+ M_i \dot f \Big)\cr
	&=: X^i \Big( f\starry i \Big)\cr}$$
where we have introduced the use of $\dot f$ for $\partial f\over\partial t$.

The action of parametric vector fields on parametric functions mimics the
action of vector fields which are orthogonal to $\dt$ in some bigger manifold,
typically $\Sigma\times\BR$, which can be thought of as a fibre bundle over
$\Sigma$.  In this interpretation, the action of $X$ on $f$ is given by taking
the horizontal lift, as specified by $\omega$.

We can similarly define parametric tensors of higher rank.
\Definition{A parametric $(p,q)$-tensor, $T\in\ptensors pq (\Sigma)$, on
$\Sigma$ is a one parameter family of $(p,q)$-tensors on $\Sigma$.  That is,
$$T:T\Sigma\cross\dots\cross T\Sigma\cross T^*\Sigma\cross\dots\cross
T^*\Sigma\cross\BR\to\BR$$
such that $T(\dots,t)\in\ptensors pq (\Sigma)$.}

As with parametric vector fields, parametric tensors can easily be expressed in
a coordinate basis
$$\generalTensor T{i_1}{i_p}{j_1}{j_q}
\fpartial{}{}x{i_1}\dots\fpartial {}{}x{i_p}\,dx^{j_1}\dots dx^{j_q}$$
where the $\generalTensor T{i_1}{i_p}{j_1}{j_q}$ are parametric functions.  We
can also talk
about 1-parameter families of metrics on $\Sigma$, that is {\it parametric
metrics}.

The Lie bracket of two vector fields orthogonal to a given family of curves
need not be a vector field orthogonal to the curves.  This ``deficiency'' is
carried over to the parametric theory, as can be seen explicitly by
calculating the action of the commutator $(XY - YX)$ on a parametric function.

$$\eqalign{X\left( Y(f)\right)&= X^i\left(Y^j f\starry j\right)\starry i\cr
	&=X^i\left(Y^j \starry i f\starry j + Y^j f\starry{ji}\right)\cr}$$
\noindent so
$$\left(XY - YX\right)(f) = \left(X^i Y^j\starry i - Y^i X^j\starry
i\right)f\starry j + X^i Y^j\left(f\starry {ji} - f\starry
{ij}\right)\Eqno$$\label\NoVF
where, in general, $f\starry {ji} - f\starry {ij}\neq 0$.

The first term on the right-hand side can indeed be written as the
(parametric) action of some vector field on $f$, but the second turns out to
involve (only) differentiation of $f$ with respect to the parameter, and hence
can not be so written.  In terms of horizontal lifts, the first term of
equation \NoVF\ is again horizontal, and can thus be identified with (the
action of) a parametric vector field, while the second term involves
differentiation in the vertical direction, which does not correspond to a
parametric vector field.

We would nevertheless like to define a notion of the ``bracket'' of parametric
vector fields.  The non-commutativity of the mixed parametric derivative makes
this non-trivial.  Without the use of a projection operator, or equivalently
of a horizontal lift, it is difficult to isolate the first term, which is the
one we want.  However, there is an intrinsic calculation that yields the
second term, which is the deficiency.  In order to define the deficiency
intrinsically we will now turn our attention to exterior differentiation of
parametric forms.

\Section{Parametric Exterior Differentiation}

Perj\'es
\Ref{{Z. Perj\'es}, {\it The Parametric Manifold Picture of Space-Time},
{Nuclear Physics} {\bf B403}, 809 (1993)} \label\PERJES
introduced a notion of exterior differentiation of parametric functions,
namely
$$d_*f = df + \omega \dot f$$
where $d$ is the usual exterior differentiation on differential forms.
Parametric functions may be considered as parametric differential 0-forms.
Parametric differential $p$-forms are just 1-parameter families of
differential $p$-forms defined on $\Sigma$.  Thus, in a coordinate basis, a
parametric differential $p$-form may be written as
$$\theta = \theta_{i_1 \dots i_p}\,dx^{i_1}\wedge\dots\wedge
dx^{i_p}$$
where the $\theta_{i_1 \dots i_p}$ are functions of $x^i$ and $t$.

There are four axioms needed to completely determine the
exterior derivative $d$ (see
\Ref{{R. Bishop and S. Goldberg}, {\bf Tensor Analysis On Manifolds}, {Dover
Publications, Inc., New York, 1980.}}\label\BandG
), namely
\item{\it i.} $df(X) = X(f)$ for functions $f$ and vector fields $X$,
\item{\it ii.} wedge-product rule: $d(\theta\wedge\tau)
		= d\theta\wedge\tau + (-1)^p \theta\wedge d\tau$
where $\theta$ is a $p$-form,
\item{\it iii.} $d(df)=0$, and
\item{\it iv.} $d$ is linear: $d(\theta + \tau) = d\theta + d\tau.$

\noindent
We already have that $d_*f(X) = X(f)$ for parametric vector fields $X$ and
parametric functions $f$.  Properties {\it ii} and {\it iv} also carry over
easily.  However, it is not clear that we wish $d_*(d_*f)=0$.  For the
parametric case, consider replacing axiom {\it iii} by
\bigskip
\item{\it iii$\,'.$} $d_*(d_*f) = 0$ for parameter-independent functions $f$.
\bigskip
Consider an exterior derivative operator, $d_*$, on parametric differential
forms satisfying {\it i}, {\it ii}, {\it iii$\,'$}, and {\it iv} for
parametric forms, vector fields, and functions.  We have the following
familiar coordinate expressions:
\item{1.} since the coordinate functions do not depend on the parameter, we
have, by {\it ii} and {\it iii}$\,'$
$$\eqalign{d_*(dx^{i_1}\dots dx^{i_p}) &= (d_*^2 x^{i_1})\wedge dx^{i_2}\,
dx^{i_3} \dots dx^{i_p}\cr
	& - dx^{i_1}\wedge (d_*^2 x^{i_2})\wedge dx^{i_3}\dots
dx^{i_p} + \dots\cr
&\dots + (-1)^{p-1} dx^{i_1}\dots dx^{i_{p-1}}\wedge (d_*^2 x^{i_p})\cr
	&= 0,}$$
\item{2.}
$$\eqalign{d_*(f dx^{i_1} \dots dx^{i_p}) &= d_*f\wedge dx^{i_i}\dots dx^{i_p}
+ f d_*(dx^{i_1}\dots dx^{i_p})\cr
	&= d_*f \wedge dx^{i_1}\dots dx^{i_p},\ {\rm and}\cr}$$
\item{3.} using {\it iv}, $d_*$ on any parametric $p$-form has the coordinate
expression
$$d_*(\theta) = d_*\left(\theta_{i_1 \dots i_p}\right)
			\wedge dx^{i_1}\dots dx^{i_p}$$
which can also be written
$$d_*(\theta) = d\theta + \omega \wedge \dot\theta.$$
It thus follows just as in the standard case that these axioms uniquely
define the {\it parametric exterior derivative} operator $d_*$.

What about $d_*(d_*f)$ on arbitrary parametric functions?  According to this
set of axioms we have
$$\eqalign{d_*^2 f &= d_*(f\starry i dx^i)\cr
	&= f \starry {ij} dx^j \wedge dx^i\cr
	&= - f\starry{ji} dx^j \wedge dx^i.\cr}$$
Therefore, $2d_*^2 f = (f\starry {ij} - f\starry {ji} )dx^j \wedge dx^i$,
which turns out to involve only parameter derivatives of $f$.  This is the
intrinsic version of the {\it deficiency}, which now measures the failure of
$d_*^2$ to be identically zero.

\Definition {The deficiency, $\D$, is the derivative operator defined by
$$\D(X,Y) f = 2 d_*^2f (X,Y),$$
for $X,Y\in\pvfs$ and $f\in\pfcns$.}

In terms of a coordinate basis we have
$$\eqalign{\D(X,Y)f &= 2d_*^2f (X^i\partiali , Y^j\partialj)\cr
	&= X^i Y^j (f\starry {ji} - f\starry {ij})\cr
	&= X^i Y^j (M_{j*i} - M_{i*j})\dot f\cr
	&=: X^iY^j\D_{ji}\dot f}$$
which is precisely the second term in \NoVF.

\Section{A Bracket Operator}

We can now easily define the bracket of two parametric vector fields
intrinsically.  We want our intrinsic definition to correspond to the
projected bracket, {\it i.e.}\ the first term of \NoVF.  But the deficiency
gives us a way to describe the second term there.  Thus, for two parametric
vector fields $X$ and $Y$, define
$$\sbracket XY f = X\left(Y(f)\right) - Y\left(X(f)\right) - \D(X,Y)f.$$
Working this out in a coordinate basis, we have
$$\eqalign{\sbracket XY f &= \left(X^i Y^j\starry i - Y^i X^j \starry
i\right)f \starry j\cr
	&+X^iY^j(f\starry{ji} - f\starry{ij}) - X^iY^j(f\starry{ji} -
f\starry{ij})\cr
	&= \left(X^i Y^j\starry i - Y^i X^j \starry
i\right)f \starry j \cr}$$
which is of course the first term in \NoVF\ as desired.
If $\lbrace x^i\rbrace$ are coordinates on $\Sigma$, then $\sbracket
{\partiali}{\partialj}=0$ as one would like.

The {\it parametric bracket operator} $\sbracket{~}{~}$ just defined fails to
satisfy the Jacobi identity, but rather satisfies a generalized (and somewhat
messy) form of this identity involving the deficiency.  However, many of the
usual properties do hold without modification.  For instance, the standard
expressions for the exterior derivatives of differential forms in terms of Lie
bracket are still valid in the parametric case.

\Theoremwoproof{If $\theta$ is a (parametric) 1-form, then
$$2\,d_*\theta(X,Y) = X\Big(\theta(Y)\Big)-Y\Big(\theta(X)\Big)
			-\theta\Big(\sbracket XY\Big)$$
for all (parametric) vector fields $X$ and $Y$.}

Given a parametric vector field $X$, we can define an $\BR$-linear mapping
$\lstar X :\pvfs\to\pvfs$ by $\lstar XY = \sbracket XY.$
Since
$$\eqalign{\lstar X(fY)g &= \sbracket X{fY} g\cr
	&= X(f)\left(Y\left(g\right)\right) - fY\left(X\left(f\right)\right) -
\D(X,fY)g\cr
	&= X(f) Y(g) + f XY(g) - fY\left(X\left(g\right)\right) - f2d_*^2g
(X,Y)\cr
	&= \left(X\left(f\right)Y + f\lstar XY\right)g\cr}$$
for all $f,g\in\pfcns$ and $X,Y\in\pvfs$, $\lstar X$ may be extended uniquely
to a parametric tensor derivation on $\Sigma$, the {\it parametric Lie
derivative}.  (See theorem 15 in Chapter 2 of
\Ref{{B. O'Neill}, {\bf Semi-Riemannian Geometry with Applications to
Relativity}, {Academic Press, Orlando, 1983}}\label\ONEILL
.)

The standard expression relating Lie differentiation, exterior
differentiation, and the interior product generalizes directly to the
parametric setting.  Specifically, letting $\ip X \alpha$ denote the obvious
extension to parametric fields of the usual interior product of a differential
form by a vector field $X$, we have the following result.

\Theorem{When acting on differential forms, parametric Lie differentiation
satisfies the operator equation
$$\lstar X = \ip X d_* + d_* \ip X$$
for any parametric vector field $X$.}
{It is straightforward to show that the right-hand side of this equation
defines a derivation.  It thus suffices to check the action of both sides on
functions and certain 1-forms.
\Footnote{It in fact suffices to check the action of both sides for parameter
independent 1-forms of the form $df$, since all parametric forms can be
written as the product of a parametric function and a parameter-independent
differential form.  It is nevertheless instructive to keep track of the
deficiency in the more general calculation presented here.}
We have
$$\eqalign{\lstar X f	&= X(f) = d_*f(X) = \ip X d_*f \cr
			&= \left(\ip X d_*f + d_* \ip X \right) f \cr
  }$$
where the last equality uses the fact that $\ip X f\equiv0$.  Furthermore,
$$\eqalign{\Big( \lstar X d_*f \Big) (Y)
	&= \lstar X \Big( d_*f(Y) \Big)
		- d_*f \left( \lstar X (Y) \right) \cr
	&= \lstar X \Big( Y(f) \Big) - d_*f \Big( \sbracket XY \Big) \cr
	&= X \Big( Y(f) \Big) - \sbracket XY (f) \cr
	&= Y \Big( X(f) \Big) + \D(X,Y) (f) \cr
	&= Y \Big( \lstar X f \Big) + 2 d_*^2 f (X,Y) \cr
	&= d_*\Big( \lstar X f \Big) (Y) + \Big( \ip X d_*^2 f \Big) (Y) .\cr
  }$$
Thus,
$$\eqalign{\lstar X d_*f &= d_*(\lstar X f) + \ip X d_*^2 f \cr
	&= d_* \ip X d_* f + \ip X d_*^2 f \cr
  }$$
and the theorem is proved.
}

\Section{Parametric Connections}

We will now introduce the notion of a connection on a parametric manifold.
Although the following definition looks identical to the definition of a
standard affine connection on a manifold, this is an illusion created by the
choice of notation.  Specifically, we have been using $X(f)$ to denote the
action of a parametric vector field on a parametric function.  The underlying
operator for such an action is {\bf not} partial differentiation, but
parametric differentiation via the operator $\partial_{*i}$.  In this sense,
one can view a parametric connection as a {\it generalized} connection on a
manifold.  \Footnote{In
\Ref{{T.~Otsuki}, Math.\ J. Okayama University {\bf 8}, 143 (1969)}
, Otsuki describes {\it generalized
connections} which do not always reduce to partial differentiation on
functions.}
That is, we generalize the notion of a vector field acting on a function.

\Definition{ An (affine) parametric connection, $\Sdel$, on $\Sigma$ is a
mapping $\Sdel : \pvfs\cross\pvfs\to\pvfs$, denoted by $\Sdel (X,Y) = \sdel
XY$,
which satisfies the following properties:
\item{\it i.} Linearity over $\pfcns: \sdel{(fX+gY)}Z = f\sdel XY + g\sdel YZ$
\item{\it ii.} Linearity:$\sdel X(Y+Z) = \sdel XY + \sdel XZ$
\item{\it iii.} Derivation: $\sdel X(fY) = X(f) Y + f\sdel XY$ for all
$X,Y,Z\in\pvfs$, $f,g\in\pfcns$, and $X(f)$ refers to the parametric action of
$X$ of $f$.}

As before, given $X\in\pvfs$ one can consider the $\BR$-linear mapping $\sdel
X :\pvfs\to\pvfs$.  Condition {\it iii} above and \ONEILL\ guarantee that
$\sdel X$ may be extended uniquely to a parametric tensor derivation on
$\Sigma$.  Thus, we may treat $\sdel X$ as a covariant derivative operator on
any parametric tensor.

We next wish to show that given a parametric metric $h$ on $\Sigma$, then there
exists a unique parametric connection on $\Sigma$ which is compatible with $h$
and torsion-free.  Hence, we need to define these last two properties.

Let $h$ be a parametric metric on $\Sigma$, denoted by $\langle\ ,\ \rangle$.
Metric
compatibility is defined in the usual way.

\Definition{A parametric connection is said to be compatible with the
parametric metric $h$ provided
$$X \Big(\metric YZ\Big) = \metric{\sdel XY}Z + \metric Y{\sdel XZ}.$$}

\Definition{The parametric torsion, $T_{\!*}$, of $\Sdel$ is defined by
$$T_*(X,Y) = \sdel XY - \sdel YX - \sbracket XY.$$
If $T_*(X,Y)=0$ for all $X,Y\in\pvfs$, then $\Sdel$ is said to be torsion
free.}

The following result generalizes to parametric connections the standard
existence and uniqueness theorem for the Levi-Civita connection.  The proof
is identical to the proof of the standard result
\Ref{{M. do Carmo}, {\bf Riemannian Geometry}, {translated by F. Flaherty,
Birkh\"auser, Boston, 1992.}}\label\DOCARMO
.{}

\Theorem{There exists a unique torsion-free parametric connection compatible
with $h$.}
{Suppose that such a $\Sdel$ exists.  Then we have
$$\eqalign{
  X\Big(\metric YZ\Big)
	&= \metric{\sdel XY}Z + \metric Y{\sdel XZ},\cr
  Y\Big(\metric ZX\Big)
	&= \metric{\sdel YZ}X + \metric Z{\sdel YX},\cr
 -Z\Big(\metric XY\Big)
	&= -\metric{\sdel ZX}Y - \metric X{\sdel ZY}.\cr}$$
Adding the above equations yields
$$\eqalign{
X\Big(\metric YZ\Big) + Y\Big(\metric ZX\Big) &- Z\Big(\metric XY\Big)\cr
	&= -\metric{\sbracket ZX}Y + \metric{\sbracket YZ}X\cr
	&+  \metric{\sbracket XY}Z + 2\metric Z{\sdel YX}.\cr
  }$$
Therefore, $\sdel YX$ is uniquely determined by
$$\eqalign{
  \metric Z{\sdel YX} &= \frac 12 \bigg(
    X\Big(\metric YZ\Big) + Y\Big(\metric ZX\Big) - Z\Big(\metric XY\Big)\cr
  &+ \metric{\sbracket ZX}Y - \metric{\sbracket YZ}X - \metric{\sbracket XY}Z
  \bigg).\cr}\Eqno$$\label\Unique
One may also use this equation to define $\Sdel$, thus proving existence.}
We can use equation \Unique\ to write out the unique parametric connection
$\Sdel$ in a coordinate basis.  If we let $\hij =
\metric{\partiali}{\partialj}$, we can define the connection symbols by
$\sdel \partiali\partialj = \gamu kij \partialk$.  Equation \Unique\ now gives
us
$$\eqalignno{\gamu lij h_{lk} &= \frac 12\left(h_{jk*i}+ h_{ki*j}-
h_{ij*k}\right)\cr
\noalign{\hbox{or}}
\gamu kij &= \frac 12 h^{km}\left(h_{jm*i} + h_{mi*j}-h_{ij*m}\right).}$$
Therefore, the connection symbols associated with $\Sdel$ agree with the
connection symbols associated with the projected covariant derivative $D$
constructed in \PaperI, which in turn agrees with Perj\'es \PERJES.


We now try to construct the curvature tensor associated with $\Sdel$.
The most obvious definition of a curvature operator would be the operator
$$S(X,Y)Z = \sdel X \sdel Y Z - \sdel Y \sdel XZ - \sdel{\sbracket XY}Z.$$
However, this turns out not to be function linear due to the fact that
$\sbracket XY f\neq XY(f) - YX(f)$.  This can, however, be easily corrected,
since we know why $S$ is not function linear (the presence of deficiency).
First, one must extend the action of $\D(X,Y)$ to tensors of rank ($p$-$q$) by
differentiating the components of an arbitrary tensor with respect to the
parameter $t$.  Since the action of $\partialt$ on $p$-forms is covariant, the
result is a ($p$-$q$) tensor.  Therefore, we define
$$Z(X,Y)W = \sdel X\sdel YW - \sdel Y \sdel XW - \sdel{\sbracket XY}W -
		\D(X,Y)W $$
and it is easily checked that this is function linear as required.
Such a definition makes use of the various derivative operators present in a
parametric theory.  Not only does the parametric manifold $\Sigma$ have the
natural parametric derivative operator $\Sdel$, but the covariant operation
of differentiation with respect to the parameter is also present, since the
deficiency operator is built out of this parametric derivative.

Given coordinates $x^i$, the components of Z may be computed as follows
$$\eqalign{
	  \zelmanovu lkij\partiall
	&= Z(\partiali , \partialj)\partialk \cr
	&= \sdel\partiali \sdel\partialj
		\partialk - \del\partialj  \sdel\partiali \partialk - 0 - 0\cr
	&= \left(\gamu ljk\starry i - \gamu lik\starry j
		+ \gamu lmi\gamu mjk - \gamu lmn\gamu mik\right) \partiall.\cr
  }$$
$Z$ is thus precisely the Zel'manov curvature reintroduced by Perj\'es
\PERJES\ and discussed in more detail in \PaperI.

\Section{Discussion}

We have shown how to recapture the projective flavor of the Gauss-Codazzi
formalism without introducing any projection operators.  After defining the
correct action of parametric vector fields on parametric functions, equation
\PVFAction , and recapturing this action in the guise of an exterior
derivative operator, the correct generalizations of Lie bracket, torsion, and
affine connection naturally followed.  Furthermore, in such an intrinsic
setting the Zel'manov curvature tensor (used by Einstein, Bergmann, Zel'manov,
and Perj\'es) is the most natural generalization of the Riemann curvature
tensor.

However, as pointed out in \PaperI, the Zel'manov curvature does {\bf not}
seem to be the natural choice in the generalized Gauss-Codazzi setting.
Rather, the Gauss-Codazzi formalism leads to the ``projected'' curvature
tensor $\rperp$.
Can one reproduce $\rperp$ intrinsically?  In terms of a coordinate basis, the
difference between $\rperp$ and $Z$ is \PaperI
$$\eqalign{\rperpu lkij - \zelmanovu lkij
	&= \left(M_{j*i} - M_{i*j}\right) h^{lm}\left(M^2 M_{m*k}
		- M^2 M_{k*m} + {\partial h_{km}\over\partial t}\right)\cr
	&= \D_{ji}h^{lm}\left(M^2\D_{mk}
		+ {\partial h_{km}\over\partial t}\right),\cr}$$
which involves both the deficiency $\D$ and the threading lapse function $M$.
As discussed in \PaperI, the appearance of $M$ is due to the presense of a
parameter $t$ whose relationship to proper ``time'' is arbitrary.  While we
have an intrinsic definition for the deficiency, we can not recover the lapse
function without explicitly introducing it.  If one is willing to add
this additional structure, then one can of course also define $\rperp$
``intrinsically'', at least in terms of its components.

Abandoning $\rperp$ for $Z$ results in a curvature operator that can be
defined entirely in terms of $\Sigma$ and the parametric structure $\omega$.
However, we know in advance that $Z$ will not possess all of the symmetries of
the Riemann curvature tensor.  In \PaperI\ it was shown that $\rperp$ is the
unique curvature satisfying Gauss' equation and, hence, enjoying all of the
inherited symmetries of the Riemann tensor (where the first Bianchi identity
for $\rperp$ resembled the identity in the presence of torsion), whereas $Z$
only enjoys some of these symmetries, namely~\PERJES
\item{\it i.} $Z(X,Y)W = -Z(Y,X)W$ and
\item{\it ii.}$Z(X,Y)W +  Z(Y,W)X + Z(W,X)Y =0$.

In the absence of deficiency, a parametric manifold can be viewed as a
1-parameter family of hypersurfaces embedded in $\Sigma\times\BR$ orthogonal
to $\omega(t)-dt$, {\it i.e.}\ such that $\omega(t)-dt$ annihilates all
vector fields tangent to the hypersurfaces.  The metric on $\Sigma\times\BR$
is not fully determined, but requires a specification of the relationship
between the parameter $t$ and arc length along the orthogonal curves, {\it
i.e.}\ the lapse function $M$.  Nevertheless, the notion of orthogonal curves
is well-defined.

Another special case is when the physical fields, including both the
parametric metric and the parametric structure, do not depend on the parameter
$t$.  In this case, the action of vector fields on (physical) functions
reduces to ordinary partial differentiation, and the parametric connection
reduces to the Levi-Civita connection of the ``parametric'' metric, which is
now a (usual) metric on the manifold of orbits.  Parametric manifolds in this
setting are thus equivalent to the formalism given by
\Ref{R.~Geroch, {\it A Method for Generating Solutions of Einstein's
Equations}, J. Math.\ Phys.\ {\bf 12}, 918 (1971).}
for spacetimes with (not necessarily hypersurface-orthogonal) Killing vectors.

But even when only the parametric structure is independent of the parameter,
in the sense that $\omega(t)$ in fact has no $t$-dependence, the structure
described here reduces to something more familiar.  Parametric exterior
differentiation can be viewed as a connection on the fibre bundle
$\Sigma\times\BR$ over $\Sigma$ precisely when the {\it horizontal subspaces}
defined by $\omega-dt$ do not depend on $t$.  This means that parametric
manifolds can be viewed as a {\it generalized fibre bundle}.  As Perj\'es has
already pointed out \PERJES, this could lead to a generalization of Yang-Mills
(gauge) theory.  Work on these issues is continuing.


\bigskip\leftline{\bf ACKNOWLEDGEMENTS}\nobreak

This work forms part of a dissertation submitted to Oregon State University
(by SB) in partial fulfillment of the requirements for the Ph.D.\ degree in
mathematics.  This work was partially funded by NSF grant PHY-9208494.

\References

\bye